\documentclass[11pt, oneside]{article}

\usepackage[T1]{fontenc}
\usepackage{arxiv}

\usepackage{amsmath}
\usepackage{amsfonts}
\usepackage{algorithmic}
\usepackage{array}
\usepackage[caption=false,font=normalsize,labelfont=sf,textfont=sf]{subfig}
\usepackage{textcomp}
\usepackage{stfloats}

\usepackage{verbatim}
\usepackage{graphicx}
\usepackage{balance}
\usepackage{xcolor}
\usepackage{todonotes}
\usepackage[normalem]{ulem}

\usepackage{booktabs, longtable, multirow, threeparttable}

\usepackage{url}
\usepackage[numbers]{natbib}
\usepackage[hidelinks]{hyperref}

\newcommand\blfootnote[1]{%
  \begingroup
  \renewcommand\thefootnote{}\footnote{#1}%
  \addtocounter{footnote}{-1}%
  \endgroup
}

\title{Opinion Dynamics with Highly Oscillating Opinions}

\author{
	Víctor A. Vargas-Pérez $^{*\text{,a,b}}$ \\
	\And
	Jesús Giráldez-Cru $^{\text{b,c}}$ \\
	\And
	Oscar Cordón $^{\text{a,b}}$ \\
}
\date{}

\begin{document}

\maketitle

\vspace{-1cm}

\begin{centering}
$^\textbf{a}$ \textit{Department of Computer Science and Artificial Intelligence (DECSAI), University of Granada (UGR), 18071 Granada, Spain}\\
$^\textbf{b}$ \textit{Andalusian Research Institute in Data Science and Computational Intelligence (DaSCI)}\\
$^\textbf{c}$ \textit{University of Seville (US), 41044 Seville, Spain} \\
\end{centering}

\blfootnote{* Corresponding Author (\textbf{\texttt{victorvp@ugr.es}})}
\blfootnote{Email addresses: \textbf{\texttt{jgiraldez@us.es}} (Jesús Giráldez-Cru), \textbf{\texttt{ocordon@decsai.ugr.es}} (Oscar Cordón)}

\begin{abstract}

Opinion Dynamics (OD) models are a particular case of Agent-Based Models in which the evolution of opinions within a population is studied. In most OD models, opinions evolve as a consequence of interactions between agents, and the opinion fusion rule defines how those opinions are updated. In consequence, despite being simplistic, OD models provide an explainable and interpretable mechanism for understanding the underlying dynamics of opinion evolution. Unfortunately, existing OD models mainly focus on explaining the evolution of (usually synthetic) opinions towards consensus, fragmentation, or polarization, but they usually fail to analyze scenarios of (real-world) highly oscillating opinions. This work overcomes this limitation by studying the ability of several OD models to reproduce highly oscillating dynamics. To this end, we formulate an optimization problem which is further solved using Evolutionary Algorithms, providing both quantitative results on the performance of the optimization and qualitative interpretations on the obtained results. Our experiments on a real-world opinion dataset about immigration from the monthly barometer of the Spanish Sociological Research Center show that the ATBCR, based on both rational and emotional mechanisms of opinion update, is the most accurate OD model for capturing highly oscillating opinions. %

\end{abstract}

\vspace{1cm}


\clearpage

%%%%%%%%%%%%%%%%%%%%%%%%%%%%%%%%%%%%%%%%%%
\section{Introduction} \label{sec:introduction}
%%%%%%%%%%%%%%%%%%%%%%%%%%%%%%%%%%%%%%%%%%

Agent-Based Models (ABM) \cite{Macal09,Epstein06} offer an appropriate framework to perform bottom-up analysis in complex systems \cite{chica2017building,Rand11,Nerrise21}. In these models, the complex dynamics of the system are inferred from the aggregation of much simpler behaviors of individual agents and their interactions. As the underlying rules governing complex systems are usually unknown, this modeling approach also provides a descriptive framework that allows us to interpret the emerging behavior of the whole system. Note that this approach is often simpler and more accurate than trying to model the entire system through global top-down rules \cite{Farmer09}.

Opinion Dynamics (OD) \cite{xia11,DongZKZL18} is a particular case of ABM \cite{FelliMMPS15} that focuses on studying how opinions spread and evolve in a population. OD models are usually classified according to (i) the representation of opinions --into discrete or continuous values--, (ii) agent interactions --into (small) fully mixed populations or (large) social networks--, and (iii) opinion fusion rule, which defines how opinions are updated after agent interactions. Representative examples of OD models, differing in these components, are the DeGroot model~\cite{Degroot74}, the Voter model~\cite{HolleyL75}, the Majority rule~\cite{Galam2002}, and OD models based on Bounded Confidence (BC)~\cite{DW00,HK02,BernardoAPV24}, among others.

\begin{figure*}[t]
    \centering
    \includegraphics[width=0.49\linewidth]{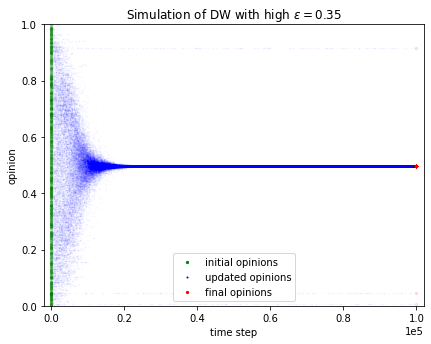}
    \includegraphics[width=0.49\linewidth]{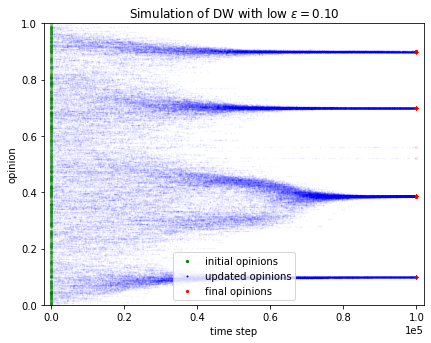}
    \caption{OD simulations on the DW model, with distinct steady states depending on the confidence threshold $\varepsilon$: consensus for high thresholds (left) and fragmentation of opinions for low thresholds (right).}
    \label{fig:example-od}
\end{figure*}

OD models are usually focused on studying the steady state reached in a simulation, i.e., the convergence towards a particular configuration of (final) opinions. In particular, three steady states are usually analyzed: (i) consensus, when all opinions have (almost) the same value, (ii) polarization, when there exist two clusters of distinct (and usually opposing) opinions, and (iii) fragmentation, when the number of distinct clusters of opinions is greater than two. Figure \ref{fig:example-od} represents an illustrative example of these phenomena on the Deffuant-Weisbuch (DW) model of BC. In this model, at every iteration a pair of agents update their opinions if they differ in less than a given confidence threshold (see Section \ref{subsec:dw} for more details on this OD model). As a result, consensus is reached with high confidence thresholds, while low values of this parameter produce a fragmentation of opinions, as showed in the figure.

In contrast, real-world opinions usually differ from these types of steady states, often exhibiting oscillating behaviors. There are many reasons that may explain the oscillations of opinions in real-world scenarios. They can be due to changes in the social conversation or because of mass media campaigns, making some topics become more or less relevant along time. In fact, many real-world opinion datasets exhibit this kind of unsteady behaviors. Interestingly enough, this real-world oscillating nature of opinions has not been studied yet in the context of OD models, to the best of our knowledge.

In this work, we study how OD models shape these oscillating patterns. To this purpose, we select three representative OD models that have both rational and emotional mechanisms of opinion update. In our analysis, we use a real-world opinion dataset about immigration, from the monthly barometer of the Spanish Sociological Research Center (in Spanish, \textit{Centro de Investigaciones Sociológicas}, CIS\footnote{\url{https://www.cis.es/catalogo-estudios/resultados-definidos/barometros}}) in the period from January 2023 to April 2024. In particular, we represent this problem as an optimization task, searching for the best parametrization of each OD model month by month. This allows us not only to measure the fitness between actual data and the best found solution, but also to provide a qualitative explanation about how the social conversation is shifting on this particular subject at every moment, with the aim of being further analyzed by experts in the field (e.g., sociologists). To solve this optimization problem, we use several Evolutionary Algorithms (EAs). Two of them are the classical Particle Swarm Optimization (PSO) \cite{pso} and Differential Evolution (DE) \cite{de}. The remaining ones are SHADE \cite{shade} and L-SHADE \cite{lshade}, which are two extensions of DE 
exhibiting good performance in practice.

Our results show that only an OD model which considers both rational and emotional mechanisms is able to explain the highly oscillating dynamics of the real scenario under study. In fact, we find that while emotional behavior tends to increase the global perceived opinion about the subject topic, rational behavior is related with a reduction in that opinion. Therefore, the lack of either of these components makes it unfeasible to fit oscillating trends such as the real-world dataset considered in our study. 

The rest of this work is organized as follows. Section~\ref{sec:related} discusses related works, while Section~\ref{sec:preliminaries} summarizes the OD models used in our study. Section~\ref{sec:framework} describes the formulation of the problem, and Section~\ref{sec:evaluation} presents the experimental results obtained. Finally, we conclude in Section~\ref{sec:conclusions}.

%%%%%%%%%%%%%%%%%%%%%%%%%%%%%%%%%%%%%%%%%%%%%%%%%%%%%%%%%%%%%%%%
\section{Related Works}
\label{sec:related}

This section briefly describes the main related works to our study. The role of memory in OD is studied in \cite{BecchettiCKPTV23}. They show that when agents have no memory (i.e., each update relies only on the current observation) the time required for a ``correct'' opinion to dominate grows quadratically with the population size. Providing each agent with just a few persistent bits accelerates that time dramatically, underscoring how even minimal recall can speed up agreement. The authors also noted that their results can be extended to settings where the population must alternate between different consensuses, considering that such oscillating behavior is fundamental to sequential decision-making processes. 

Binary OD with biased influence towards one option is analyzed in \cite{Anagnostopoulos20}. Under a biased majority rule, consensus can take extraordinarily long on dense networks, whereas a biased voter rule converges much faster, mirroring the pace of repeated random sampling. This picture was refined in \cite{LesfariGP22} by tying slowdowns to specific graph structures. They prove that certain sparse but well-organized networks still exhibit extremely long convergence times, while more loosely connected cubic graphs do not, revealing a sharp topology-driven transition.

The effects of local interactions in OD are examined in \cite{FotakisPS16}. They extended the Friedkin-Johnsen (FJ) and Hegselmann-Krause (HK) models to settings where each agent consults only a limited subset of neighbors. With such partial information, the classical FJ process slows down dramatically, and the network-based HK variant forms more opinion clusters. However, allowing each agent to poll just a logarithmic number of neighbors almost restores the speed and clustering of the fully informed model. These results quantify the trade-off between communication cost and the time to reach a stable state. 

An analysis on sampling public opinions considering the correlation of the social network is presented in \cite{HuangLC17}. They propose an optimal partitioned sampling scheme that groups similar individuals and surveys one representative per group, reducing estimation error well below that of simple random sampling. A fast greedy algorithm captures most of these gains and remains effective even when similarity estimates are noisy, as confirmed on both synthetic graphs and real data. 

Across these lines of research, the common endpoint is a stable consensus or at least a fixed set of opinion clusters. Unfortunately, none of these works explicitly study OD in the context of real-world highly oscillating opinions. Our study targets this overlooked regime. 

%%%%%%%%%%%%%%%%%%%%%%%%%%%%%%%%%%%%%%%%%%%%%%%%%%%%%%%%%%%%%%%%
\section{Preliminaries}
\label{sec:preliminaries}

Let us consider a population of $N$ agents interacting in a social network, represented by a graph $G(V,A)$, with $V$ being the set of nodes (with $|V|=N$), and $A$ being its adjacency matrix. Each agent is represented by a node $i$ in the graph, and has social interactions with other agents connected to it, i.e., agents $j$ such that $A_{ij}=1$. Therefore, this graph represents the scope of the social conversation every agent has.

For the sake of simplicity, let us consider that agents only have one opinion on one particular subject.\footnote{The extension to multiple opinions/subjects is straightforward.} This opinion evolves due to social interactions during a number of time steps $T$. Let $x_i(t) \in [0,1]$ be the opinion of agent $i$ at time step $t$, with $1\leq i \leq N$ and $0 \leq t \leq T$. The opinion profile $X(t) = \{x_i(t)\}_{1 \leq i \leq N}$ represents the set of agents' opinions at time step $t$. As a consequence, OD models study the transition from $X(0)$ to $X(T)$, and the opinion state achieved in $X(T)$. In what follows, we briefly summarize the opinion fusion rule of the OD models analyzed in this work, which is used to update $x_i(t)$.

\subsection{The Friedkin-Johnsen Model}

The Friedkin-Johnsen (FJ) model \cite{FJmodel} generalizes the DeGroot model \cite{Degroot74} considering that agents are adhered to their original opinion to a certain degree. The opinion fusion rule in the FJ model is defined as:
\begin{equation} \label{eq:fj}
    x_i(t+1) = \xi_i \sum_{j=1}^{N}{\left ( w_{ij}x_j(t) \right )} + (1-\xi_i)x_i(0)
\end{equation}
\noindent 
where $\xi_i$ is the susceptibility of agent $i$, and $w_{ij}$ is the weight agent $i$ gives to agent $j$. Notice that $(1-\xi_i)$ represents the extent of adherence (stubbornness) to its original opinion $x_i(0)$.

\subsection{The Deffuant-Weisbuch Model}
\label{subsec:dw}

The DW model \cite{DW00} is a model of BC, i.e., agents are only influenced by other agents that belong to their confidence area. This confidence area is defined by a confidence threshold $\varepsilon$, and based on its value, this model is able to explain both consensus and fragmentation of opinions: high values of $\varepsilon$ for the former, and low values for the latter (as showed in Figure \ref{fig:example-od}). The opinion fusion rule is defined as:
\begin{align}
    \begin{split}
        x_i(t+1) = x_i(t) &+ \mu(x_j(t) - x_i(t)) \\
            &\textrm{\,\,iff\,\,} |x_i(t) - x_j(t)| < \varepsilon 
    \end{split}
    \label{eq:dw}
\end{align}
\noindent
where $\mu \in [0,0.5]$ is the convergence speed.

\subsection{The ATBCR Model}

The Agent-independent Time-based Bounded Confidence and Repulsion (ATBCR) model \cite{ATBCR} introduces into DW an extension based on repulsion behaviors. The purpose of this new rule is to have a mechanism to explain not only consensus and fragmentation, but also extremization (i.e., opinions with extreme values). The opinion fusion rule completes Equation~\ref{eq:dw} by also introducing the following rule:
\begin{align}
    \begin{split}
        x_i(t+1) = x_i(t) &- \mu(x_j(t) - x_i(t))\\
            &\textrm{\,\,iff\,\,} |x_i(t) - x_j(t)| > \vartheta_i(t) 
    \end{split}
    \label{eq:atbcr}
\end{align}
\noindent
where $\vartheta(t)$ is the polarization threshold, and the other parameters are as in the DW model. Values outside the interval $[0,1]$ are truncated to the extreme values $0$ or $1$.

Notice that this model uses both confidence and polarization thresholds ($\varepsilon_i(t)$ and $\vartheta_i(t)$) that are specific for each agent $i$ and time step $t$. However, for comparison purposes, in this work we will only use the agent-uniform version of ATBCR, which uses, as in the DW model, global thresholds for the whole population, referred from now on as $\varepsilon(t)$ and $\vartheta(t)$. This adaptation, as well as the ones in the FJ and DW models, are precisely described in Section \ref{subsec:parameters}.

\subsection{Rationale of the Selected OD Models}

The FJ model is a purely emotional system, since agents are attached to their original opinion. In contrast, DW represents a purely rational system where OD are only driven by a BC mechanism. Finally, ATBCR is a hybrid method, with both rational and emotional mechanisms, since it extends DW by incorporating a polarization rule, which is usually considered as an emotional behavior.

%%%%%%%%%%%%%%%%%%%%%%%%%%%%%%%%%%%%%%%%%%%%%%%%%%%%%%%%%%%%%%%%
\section{Proposed Methodology}
\label{sec:framework}

This section outlines the methodology employed to extend and study the OD models described in Section \ref{sec:preliminaries} for real-world scenarios with highly oscillating opinions. First, we introduce dynamic parameters to existing OD models. Second, we describe the synthetic social network used. Third, we detail how we leverage the CIS monthly barometer data to initialize agent population. We then formulate the optimization problem to be addressed. Finally, we discuss the fitness evaluation of model configurations against real-world data. 

\subsection{Dynamic Parameters in OD Models}
\label{subsec:parameters}

Let us consider a time series $P = [p_1, \dots , p_k]$ representing the periods of the study. In our case, it comprises $k=15$ monthly periods from January 2023 to April 2024. Traditional FJ and DW models assume static parameters, where $\xi$ and $\varepsilon$ remain fixed for all agents and for all time steps. We extend these models to incorporate period-dynamic parameters, $\xi(p_i)$ and $\varepsilon(p_i)$, respectively, alongside a period-dependent convergence speed $\mu(p_i)$ for DW and ATBCR models (for $p_i \in P$). Similarly, we use a uniform version of the ATBCR model with period-dependent parameters $\varepsilon(p_i)$ and $\vartheta(p_i)$, which are shared by all the agents in the population. In our problem, each period $p_i \in P$ consists of many time steps (representing agent's interactions), and the values of the OD parameters remain fixed during the whole period. These dynamic parameters allow us to change the behavior of the OD models over time, essential for capturing the oscillating opinion trends observed in real-world scenarios.

\subsection{Agents Interactions through a Social Network}

In our study, we constrain the social interactions of agents to a synthetic scale-free social network generated by the Barabási-Albert (BA) preferential attachment algorithm \cite{barabasi}. This algorithm produces networks with a power-law degree distribution, where most nodes have few connections, and highly connected hubs facilitate rapid information flow. Such properties align with those observed in real-world social networks \cite{barabasi}, making the BA network suitable for our study. We set the parameter $m=3$ (number of edges added per iteration of the BA algorithm), resulting in a network with an average degree of 6. The number of agents $N$ (i.e., nodes in the social network) corresponds to the number of respondents of the January 2023 CIS barometer, which is $N$ = 3,961.

In practice, the use of the social network implies that in the FJ model, $w_{ij} = A_{ij}/\text{deg}(i)$, where $\text{deg}(i)$ is the degree of node $i$. In the DW and ATBCR models, the fusion rule (Equations \ref{eq:dw} and \ref{eq:atbcr}) can be applied only to pairs of agents that are connected, i.e., agents $i$ and $j$ such that $A_{ij}=1$.

\subsection{Initialization of the Population with Real-World Survey Data} \label{sec:initialization}
\label{subsec:init}

In this study, we define the opinion of each agent, $x_i(t) \in [0, 1]$, as the level of concern of that agent regarding the immigration topic. The Spanish Sociological Research Center (CIS) conducts a monthly barometer survey to collect opinions from the Spanish population on various topics. One of these periodic questions is ``\textit{In your opinion, what is the main problem currently facing Spain? And the second? And the third?}''. This is the only monthly question in these surveys from which we can derive an opinion profile of the Spanish population regarding immigration. Specifically, we can determine whether each respondent includes this topic in their answers and in what order of importance.

However, the resulting dataset provides categorical values, so we need to translate the answers into the continuous domain $[0,1]$. To achieve this, we define a concern threshold parameter $c_{\text{th}} \in [0,1]$. An agent $i$ is considered concerned or not at time step $t$ depending on whether their opinion $x_i(t) \geq c_{\text{th}}$. 

Using this approach, we initialize the population of agents based on CIS barometer data from January 2023, the first month of the study period. We perform an 1:1 mapping from real survey respondents to agents and generate the initial opinion $x_i(0)$ of each agent by sampling a random value from a Gaussian distribution $\mathcal{N}(\mu,\sigma)$, where $\mu$ is the mean and $\sigma$ is the standard deviation.\footnote{Here, $\mu$ and $\sigma$ adhere to the standard notation for mean and standard deviation used in Gaussian distributions. Note that this $\mu$, only used along Section \ref{subsec:init}, should not be mistaken for the convergence speed parameter used in the DW and ATBCR models, as defined in Section \ref{sec:preliminaries} and used in the rest of this work.} Initializing agent population parameters through statistical distributions is a common practice in ABM literature \cite{rand21}. In our case, the parameters of the distribution depend on respondent's answer. Specifically, we define the following cases:

\begin{itemize}
    \item If the respondent does not mention immigration in their answers, we set $x_i(0) \sim \mathcal{N}(\frac{c_{th}}{2}, \frac{c_{th}}{6})$.
    \item If the respondent identifies immigration as the $k$-th most important problem ($k \in \{1,2,3\}$), we set $x_i(0) \sim \mathcal{N}\left(c_{\text{th}} + \frac{7-2k}{6} \cdot (1 - c_{\text{th}}), \frac{(1 - c_{\text{th}})}{18}\right)$. 
\end{itemize}

\begin{figure}
  \includegraphics[width=\linewidth]{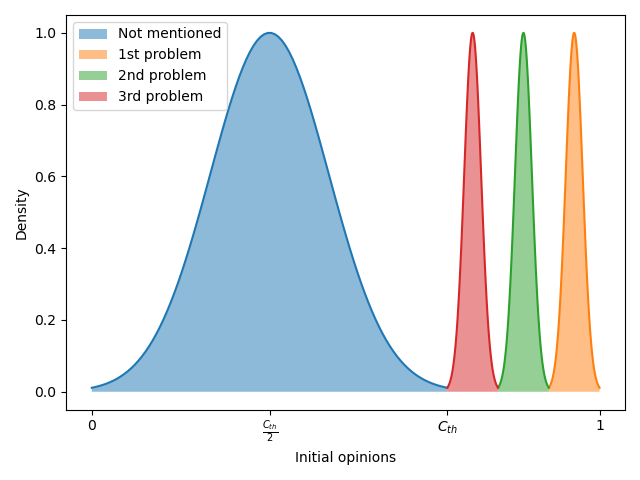}
  \caption{Illustration of Gaussian distributions for initializing agents' opinions based on survey answers. The concern threshold $c_{\text{th}}=0.7$ defines the minimum opinion value for concern in this example. Upper-half distributions share the same standard deviation and divide $[c_{\text{th}}, 1]$ into three equal intervals.}
  \label{fig:initial_opinions}
\end{figure}

The idea behind these distributions is illustrated in Figure \ref{fig:initial_opinions}. This approach ensures that the generated values cover the range $[0,1]$, with the concern threshold $c_{\text{th}}$ dividing the range into two regions: one for respondents who are not concerned and another for those who are. The upper half of the range is further split into three equal segments, each corresponding to the importance assigned to immigration in the respondent's answers. The majority of values in a Gaussian distribution ($99.73\%$) lie within the interval $[\mu - 3\sigma, \mu + 3\sigma]$. Therefore, the standard deviation of each distribution is carefully selected following this rule to ensure that the sampled values cover the intended range. Additionally, to prevent outlier values, the sampled values are truncated to their corresponding interval. For example, if a respondent is not concerned, the generated value is constrained to the range $[0, c_{\text{th}}]$.

\subsection{Problem Formulation}
\label{subsec:formulation}

Let us consider an OD model $OD(\alpha_1, \dots, \alpha_n)$ characterized by $n$ free parameters (e.g., confidence threshold, convergence speed, etc.). Let $s(p_i)$ be the size (i.e., number of time steps) of period $p_i \in P$, $e(p_i) = \sum_{j=1}^{i}{s(p_j)}$ the time step when period $p_i$ ends. Let $c(p_i)$ denote the proportion of concerned agents at the end of period $p_i \in P$, i.e., $c(p_i) = |\{x_j(e(p_i)) \geq c_{\text{th}}\}|$. Similarly, let $h(p_i)$ be the actual proportion of concerned individuals in the historical data in the same period $p_i$.

For a given fitness function $F$ (discussed in more details in Section \ref{subsec:fitness}), let $F(c(p_i), h(p_i))$ be the error of period $p_i \in P$, i.e., how well the OD model, with a particular parametrization $\{\alpha_1, \dots , \alpha_n\}$, fits historical data; and the average error $\bar{F}$ as:
\begin{equation}
    \bar{F} = \frac{1}{|P|}\sum_{p_i \in P} F(c(p_i), h(p_i))
\end{equation}

Our optimization problem consists of finding the best parametrization $\{\alpha_1, \dots , \alpha_n\}$ such that $\bar{F}$ is minimized, subject to parameter domains (e.g., $\mu(p_i) \in [0,.5]$), and other qualitative constraints that may be added, e.g., the confidence and the polarization thresholds do not collapse into the same value (in particular, we use $0.1 \leq \vartheta(p_i) - \varepsilon(p_i)$). Precise details on these constraints are presented in Section \ref{sec:evaluation}.

\subsection{Model Evaluation} \label{sec:fitness}
\label{subsec:fitness}

We define a fitness function that quantifies the similarity between the model output and historical data across different time steps. Specifically, we use the Mean Absolute Percentage Error (MAPE) of the proportion of concerned agents as the fitness function. MAPE is a recommended metric for model calibration, particularly when the goal is to capture general dynamics trends rather than specific data matches \cite{chica2017building}, as in the case in our study. The MAPE fitness function is defined as:

\begin{equation}
    \text{MAPE} = \frac{100}{|P|} \sum_{p_i \in P}{\frac{|c(p_i) - h(p_i)|}{h(p_i)}}
    \label{eq:fitness}
\end{equation}

The lower the MAPE value, the better the model fits the historical data. Since the OD models are stochastic, we perform 20 independent Monte Carlo simulation runs for each evaluation.\footnote{This number was determined based on a preliminary analysis showing that the average final opinion of the population stabilizes across different simulations after 20 runs.} The final fitness value is the average of the MAPE values obtained from these 20 independent runs.  

In the OD models, each time step corresponds to a single social interaction, with each real day equivalent to 450 time steps.\footnote{Previous studies used these OD models with 45 time steps per day \cite{giraldez2024modeling}. Here, we scale this value by a factor of 10 since immigration, as a general public opinion topic, is discussed more frequently than more specific subjects like superstars' opinions. In any case, this choice is not critical, as its effect is balanced by the convergence speed (ATBCR and DW models) and susceptibility level (FJ model), both free parameters that we calibrate.} Given that a month consists of approximately 30 days, each period $p \in P$ comprises approximately 13,500 time steps. The simulation period spans from January 2023 to April 2024, with an available barometer survey for each month,\footnote{Except August 2023. However, we will also consider this month to make equal the length of all periods, although it does not participate in the fitness function.} resulting in $|P|=15$ evaluation points for the fitness function.

%%%%%%%%%%%%%%%%%%%%%%%%%%%%%%%%%%%%%%%%%%%%%%%%%%%%%%%%%%%%%%%%
\section{Experimental Evaluation}
\label{sec:evaluation}

The goal of our experiments is to identify the best OD model configuration to represent the oscillating historical population concern given by CIS barometer data from January 2023 to April 2024. To achieve this, we define in Section \ref{sec:setup} an experimental setup that considers different problem variants derived from the concern threshold $c_{\text{th}}$ and the OD models, as well as various EAs to solve them. Section \ref{sec:results} presents and discusses the results obtained from these experiments.

\subsection{Experimental Setup} \label{sec:setup}

We define an optimization problem for each OD model, where the objective is to find their dynamic parameters, as defined in Section \ref{subsec:formulation}, minimizing the fitness function outlined in Section \ref{sec:fitness}. One value is assigned per month and parameter, resulting in 15 $\times$ 1 = 15 parameters for the FJ model, 15 $\times$ 2 = 30 parameters for the DW model, and 15 $\times$ 3 = 45 parameters for the ATBCR model. Table \ref{tab:problems} summarizes the specific monthly parameters for each OD model and the constraints considered. These constraints ensure that the calibrated parameters remain within a reasonable range, avoiding extreme values that could result in unrealistic behaviors.

\begin{table}[!htb]
  \centering
  \caption{Monthly parameters and constraints for each OD model. 
  }
  \label{tab:problems}
  \begin{tabular}{@{}llc@{}}
  \toprule
  \textbf{OD Model} & \textbf{Parameters by month} & \textbf{Constraints}             \\ \midrule
  \textbf{FJ}       & Susceptibility $\xi(p)$                       & $\xi(p) \in$ {[}0.1, 1{]} \\ \midrule
  \textbf{DW} & \begin{tabular}[c]{@{}l@{}}Convergence speed $\mu(p)$\\ Confidence threshold $\varepsilon(p)$\end{tabular}
     &
    \begin{tabular}[c]{@{}l@{}}$\mu(p) \in$ (0, 0.5{]}\\ $\varepsilon(p) \in$ {[}0, 0.5{]}\end{tabular} \\ \midrule
  \textbf{ATBCR} & \begin{tabular}[c]{@{}l@{}}Convergence speed $\mu(p)$\\ Confidence threshold $\varepsilon(p)$\\ Polarization threshold $\vartheta(p)$\end{tabular}
     &
    \begin{tabular}[c]{@{}c@{}}$\mu(p) \in$ (0, 0.5{]}\\ $\varepsilon(p) \in$ {[}0, 0.5{]}\\ $\vartheta(p) \in$ {[}0.5, 1.0{]}\\ 0.1 $\leq \vartheta(p) - \varepsilon(p) \leq$ 0.9\end{tabular} \\ \bottomrule
  \end{tabular}
  \end{table}

Given the dimensionality of these optimization problems, we use EAs to solve them. Previous studies have demonstrated the effectiveness of these algorithms for calibrating ABMs \cite{malleson2014optimising,nguyen2014review,zhong2015differential}. We consider four EAs: the classic DE (specifically, the DE/rand/1/bin variant), the Success-History based DE variant (SHADE), its extension with linear population reduction (L-SHADE), and PSO. Table \ref{tab:hyperparameters} shows the hyperparameter values used for each EA. According to \cite{de}, SF = 0.5 is an effective initial choice for DE. Similarly, the control memory size H originally proposed for SHADE was 100 \cite{shade}, while the same authors reduced this value to 6 for L-SHADE \cite{lshade}. For PSO, we adopt the general hyperparameter configuration recommended by \cite{pso}, which, combined with the velocity constriction to each variable's range, results in a PSO with no problem-specific parameters. Finally, to ensure a fair comparison, we set a population size of 100 solutions for all algorithms\footnote{This is the initial size for L-SHADE, whose population decreases linearly in each generation.} and a maximum number of 30,000 evaluations as the stopping criterion.

\begin{table}[!htb]
\centering
\caption{Hyperparameters values used for each EA.}
\label{tab:hyperparameters}

\begin{tabular}{@{}lll@{}}
\toprule
\textbf{EA}      & \textbf{Hyperparameter} & \textbf{Value} \\ \midrule
\textbf{DE}      & Crossover rate (CR)     & 0.5            \\
                 & Scaling factor (SF)      & 0.5            \\ \midrule
\textbf{SHADE}   & Control memory size (H) & 100            \\ \midrule
\textbf{L-SHADE} & Control memory size (H) & 6              \\ \midrule
\textbf{PSO}     & Social weight (C1)      & 1.49618        \\
\textbf{}        & Cognitive weight (C2)   & 1.49618        \\
                 & Inertia weight (W)      & 0.7298         \\ \midrule
\textbf{Common}   & Population size & 100            \\ 
\textbf{hyperparameters}   & Number of evaluations & 30,000            \\ 
                 \bottomrule
\end{tabular}

\end{table}

To fully establish this optimization problem, we need to set the concern threshold $c_{\text{th}}$, as explained in Section \ref{sec:initialization}. We consider three values: $c_{\text{th}} = \{0.60, 0.75, 0.90\}$, as representative values for this task. Therefore, this results in 3 $\times$ 3 = 9 optimization problems and 4 EAs to solve them, leading to 36 calibration tasks in total. 

We perform the experiments on a server with 2 $\times$ 2.2 GHz Intel Xeon Silver 4214 (12 cores), 256 GB DDR4 RAM. This setup allows us for parallel execution of the 20 Monte Carlo simulations involved in each fitness evaluation. The 30,000 evaluations required for each calibration task take approximately 96 hours to complete.

\subsection{Results and Discussion} \label{sec:results}

We present the results obtained for the 36 
calibrations in Table \ref{tab:mape}. From these results, it can be observed that the OD model with the lowest fitness value is ATBCR in all cases, followed by DW and FJ. Additionally, it is noted that the ATBCR model yields better results as the confidence threshold $c_{\text{th}}$ increases. However, this behavior is not observed for the FJ and DW models. In fact, while the DW model shows consistent results across different confidence thresholds, the FJ model exhibits an opposite pattern to ATBCR, with performance deteriorating as a higher threshold is considered.

\begin{figure}
  \centering
  \includegraphics[width=0.9\linewidth]{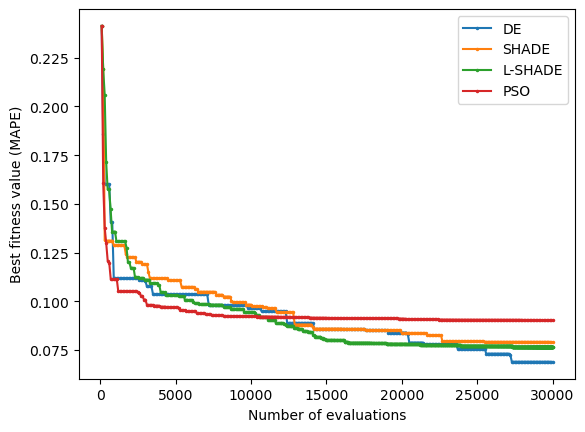}
  \caption{Fitness evolution of all optimization algorithms for the ATBCR model in the scenario with a concern threshold of 0.9.}
  \label{fig:fitness_evolution}
\end{figure}

Regarding the differences between the various EAs, they are not particularly pronounced. The general pattern indicates that PSO consistently yields the worst results, while the best solutions are always obtained with DE or L-SHADE, with SHADE performing similarly to these two. Notably, the most significant differences between EAs are observed with the ATBCR model, whereas the results for the other models are more homogeneous. This could suggest that the ATBCR model's parametrization is more flexible and, consequently, more sensitive to the choice of optimization algorithm.

\begin{table*}[!htb]
  \centering
  \caption{Fitness value (MAPE) of the solution obtained by each EA for all concern scenarios and OD models. The minimum value for each concern threshold is highlighted in bold. The best solution found is underline.}
  \label{tab:mape}
  \resizebox{\textwidth}{!}{
  \begin{tabular}{@{}lrrr|rrr|rrr@{}}
  \toprule
  \textbf{Algorithm} &
    \multicolumn{3}{c|}{\textbf{Concern $c_{\text{th}} = 0.60$}} &
    \multicolumn{3}{c|}{\textbf{Concern $c_{\text{th}} = 0.75$}} &
    \multicolumn{3}{c}{\textbf{Concern $c_{\text{th}} = 0.90$}} \\ \midrule
   &
    \multicolumn{1}{c}{\textbf{FJ}} &
    \multicolumn{1}{c}{\textbf{DW}} &
    \multicolumn{1}{c|}{\textbf{ATBCR}} &
    \multicolumn{1}{c}{\textbf{FJ}} &
    \multicolumn{1}{c}{\textbf{DW}} &
    \multicolumn{1}{c|}{\textbf{ATBCR}} &
    \multicolumn{1}{c}{\textbf{FJ}} &
    \multicolumn{1}{c}{\textbf{DW}} &
    \multicolumn{1}{c}{\textbf{ATBCR}} \\ \cmidrule(l){2-10} 
  \textbf{DE} &
    3.61e-01 &
    3.55e-01 &
    1.87e-01 &
    4.61e-01 &
    3.60e-01 &
    8.80e-02 &
    7.11e-01 &
    3.60e-01 &
    \textbf{\uline{6.89e-02}} \\
  \textbf{SHADE} &
    3.61e-01 &
    3.56e-01 &
    1.81e-01 &
    4.61e-01 &
    3.60e-01 &
    8.61e-02 &
    7.11e-01 &
    3.60e-01 &
    7.91e-02 \\
  \textbf{L-SHADE} &
    3.61e-01 &
    3.55e-01 &
    \textbf{1.51e-01} &
    4.61e-01 &
    3.60e-01 &
    \textbf{8.51e-02} &
    7.11e-01 &
    3.60e-01 &
    7.65e-02 \\
  \textbf{PSO} &
    4.03e-01 &
    3.56e-01 &
    2.32e-01 &
    5.64e-01 &
    3.60e-01 &
    9.97e-02 &
    7.82e-01 &
    3.61e-01 &
    9.04e-02 \\ \bottomrule
  \end{tabular}
  }
  \end{table*}

  \begin{figure*}[!htb]
    \centering
    \includegraphics[width=0.49\textwidth]{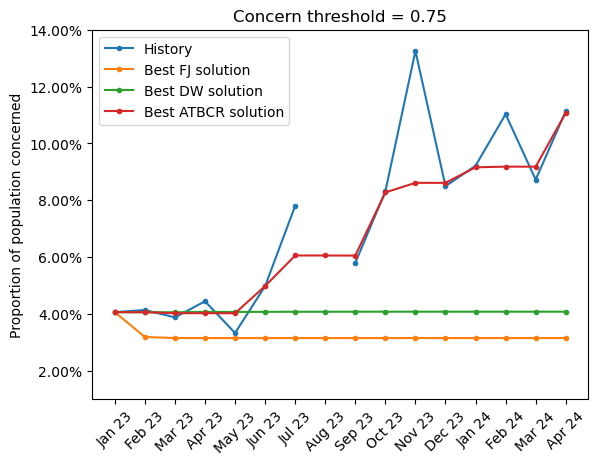}
    \includegraphics[width=0.49\textwidth]{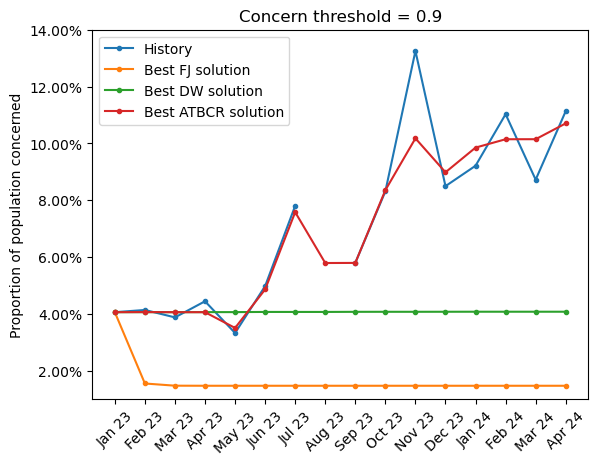}
    \caption{Comparison of the proportion of concerned population in historical data (blue) with respect to the best solution found for each OD model in the concern threshold scenarios 0.75 and 0.9.}
    \label{fig:concern_evolution}
\end{figure*}

Figure \ref{fig:fitness_evolution} shows the evolution of the best fitness value achieved by each EA for the problem defined by the ATBCR model with $c_{\text{th}} = 0.9$. PSO improves rapidly during the initial evaluations but stagnates quickly after approximately 5,000 evaluations. In contrast, the DE variants improve more gradually but outperform PSO at around 10,200 evaluations. L-SHADE reaches better results quickly. However, in this problem, basic DE continues to improve almost until the end of the calibration process, ultimately achieving the best solution with a MAPE of 6.89e-02.

To analyze the results in more detail, Figure \ref{fig:concern_evolution} shows the evolution of the proportion of concerned population in the historical data (in blue) compared to the best solutions found for each OD model under the concern threshold scenarios of 0.75 and 0.9. These figures provide a clearer explanation of the differences in MAPE values observed in Table \ref{tab:mape}. For the DW model, the solution leads to an almost constant concern at the initial value, while for FJ it is also almost constant but deviating from the initial value. We find that the best parameterizations for DW effectively deactivate the mechanism, assigning a confidence threshold close to 0 for all months. Similarly, the best solutions for the FJ model yield a susceptibility of 0.1 across all months. This value corresponds to the minimum allowed (see Table \ref{tab:problems}), and a null susceptibility would equate to maintaining the initial opinion unchanged (see Equation \ref{eq:fj}). Thus, the EAs identify the best solutions for DW and FJ as those that nearly deactivate the diffusion mechanism entirely for all months. This suggests that these OD models lack the capability to capture the evolution of concern from the real data, even when their parameters are allowed to vary over time.

In contrast, the results found with ATBCR are positive. With $c_{\text{th}} = 0.75$, the model demonstrates a reasonable evolution, effectively capturing the increases in the historical data in a staggered manner. The best solution among the 36 calibrations is achieved by DE applied to the ATBCR model with $c_{\text{th}} = 0.9$, a result that is also visually evident in Figure \ref{fig:concern_evolution}. In this scenario, the model successfully captures the moments of decreasing concern, producing results that closely match the historical data.

\begin{table}[!htb]
  \centering
  \caption{Parameters of the best found solution (obtained by DE for ATBCR with $c_{\text{th}}=0.90$).}
  \label{tab:bestsolution}
  \begin{tabular}{@{}rccc@{}}
  \toprule
  \textbf{Month} &
    \textbf{\begin{tabular}[c]{@{}l@{}}Convergence\\ Speed $\mu(p)$\end{tabular}} &
    \textbf{\begin{tabular}[c]{@{}l@{}}Confidence\\ Threshold $\varepsilon(p)$\end{tabular}} &
    \textbf{\begin{tabular}[c]{@{}lr@{}}Polarization\\ Threshold $\vartheta(p)$\end{tabular}} \\ \midrule
  \textbf{January 2023}   & 0.15 & 0.07 & 0.85 \\
  \textbf{February 2023}  & 0.01 & 0.27 & 0.65 \\
  \textbf{March 2023}     & 0.01 & 0.11 & 1.00 \\
  \textbf{April 2023}     & 0.32 & 0.30 & 0.99 \\
  \textbf{May 2023}       & 0.30 & 0.02 & 0.50 \\
  \textbf{June 2023}      & 0.43 & 0.01 & 0.54 \\
  \textbf{July 2023}      & 0.15 & 0.50 & 1.00 \\
  \textbf{August 2023}    & 0.01 & 0.00 & 0.86 \\
  \textbf{September 2023} & 0.49 & 0.02 & 0.50 \\
  \textbf{October 2023}   & 0.39 & 0.02 & 0.50 \\
  \textbf{November 2023}  & 0.29 & 0.50 & 0.97 \\
  \textbf{December 2023}  & 0.25 & 0.00 & 0.64 \\
  \textbf{January 2024}   & 0.36 & 0.19 & 0.73 \\
  \textbf{February 2024}  & 0.50 & 0.06 & 0.96 \\
  \textbf{March 2024}     & 0.50 & 0.03 & 0.52 \\ \bottomrule
  \end{tabular}
  \end{table}

Having developed a model capable of capturing the highly oscillating dynamics of concern from real-world data, we can qualitatively analyze its parametrization to understand how social conversation evolves on this specific topic over time. Table \ref{tab:bestsolution} presents the parameters of this solution. Note 
that 
the effect of the parameters for each month is reflected in the concern levels observed in the subsequent month (see Fig. \ref{fig:concern_evolution}).

It is important to recall that higher confidence (rational mechanism) corresponds to a high confidence threshold, while greater polarization (emotional mechanism) corresponds to a low polarization threshold. The key finding is that periods of increasing concern are associated with heightened polarization and reduced confidence (e.g., May, June, and September 2023). Conversely, periods of decreasing concern are linked to lower polarization and increased confidence (e.g., July and November 2023). Additionally, the convergence speed acts as a regulatory parameter that generally increases when a faster shift in opinions is required. For instance, during June and September, values close to 0.5 indicate a need for quicker opinion changes.

In consequence, our model is not only able to find a solution close to the optimal one, but also to provide qualitative results on the solution found, which can be further examined by domain experts to better understand the reasons that shift social conversation and opinion formation.

%%%%%%%%%%%%%%%%%%%%%%%%%%%%%%%%%%%%%%%%%%%%%%%%%%%%%%%%%%%%%%%%
\section{Concluding Remarks}
\label{sec:conclusions}

In this work, we have studied the ability of different OD models to fit and explain the highly oscillatory dynamics of opinions in a real-world scenario. To achieve this, we considered versions of these OD models with dynamic temporal parameters, allowing them to change their behavior in line with shifts in opinion trends over time. This approach results in a significant increase in the number of parameters (proportional to the number of temporal changes considered), making it necessary to carry out an optimization process using heuristic search techniques such as EAs.

Our results show that the highly oscillating opinions in the studied scenario can only be explained by an OD model that incorporates both emotional and rational mechanisms, as is the case with ATBCR. Simpler models focused solely on one of these mechanisms, such as FJ and DW, are unable to account for the observed evolution of opinions.

The best solution found with the ATBCR model validates the success of our methodology, achieving a result very close to the historical data of the real scenario studied, with a MAPE of 6.89e-02. This level of fit supports the reliability of the model in explaining the OD of the real scenario. In our case, we studied the monthly evolution of concern among the Spanish population about immigration. Based on our solution, we observed that periods of increased concern stem from more emotional behavior, with a population more prone to polarization. Conversely, decreases in concern correspond to a more rational population, with a greater ability to reach consensus. 

As future work, we propose conducting a more in-depth qualitative study of the solution found, with 
domain expert assistance. This would allow us, for example, to identify the events or occurrences during the studied period that explain the population’s behavioral changes each month. Additionally, we propose designing what-if scenarios to test the model's response to changes in agent behavior due to, e.g., the impact of hypothetical public awareness campaigns.

Moreover, the methodology employed in this work is sufficiently generic to extend beyond its application to the study of public opinion on immigration. The same approach could be directly applied to other public opinion topics, such as public health or climate change.

A limitation of our work is the nature of our data. The surveys we used do not directly ask about the concern for studied topic, but rather inquire about which are the three main problems of the country. This requires us to relate these issues by assuming that if a respondent mentions the studied topic, their concern level must be high, and we must choose a threshold to make this distinction. This opinion initialization, although realistic, may hinder some unknown patterns of opinion evolution. Therefore, we encourage the collection of data on the studied topic with a more straightforward translation to an opinion value. However, it is crucial to still maintain a high temporal resolution (monthly in our case) to ensure a reliable historical dataset that supports the confidence in the OD models that fit it. 

%%%%%%%%%%%%%%%%%%%%%%%%%%%%%  BEGIN ACKNOWLEDGMENTS   %%%%%%%%%%%%%%%%%%%%%%%%%%%%%

\section*{Acknowledgments} 

This work was supported in part by MCIN/AEI/10.13039/501100011033 and ERDF “A way of making Europe” under Grant CONFIA PID2021-122916NB-I00, in part by the FPU Program under Grant FPU20/02441, in part by Grant RYC2022-036395-I funded by MICIU/AEI/10.13039/501100011033 and ESF+, and in part by the ``Plan Propio de Investigación y Transferencia de la Universidad de Granada'' under grant PPJIB2023-048.

\end{document}